\pdfoutput=1
\documentclass[11pt,twoside,a4paper,cmspaper,final,collab]{cms-tdr}
\def\svnVersion{2e5aaf7}\def\svnDate{2025/06/26}\def\cmsCernNoTag{CERN-EP-2024-135}\def\cmsCernDate{\today}\def\cmsMessage{Published in Physics Letters B as \href{http://dx.doi.org/10.1016/j.physletb.2025.139672}{\doi{10.1016/j.physletb.2025.139672}.}}
\begin{document}\cmsNoteHeader{HIN-21-016}

\newcommand{\noff}     {\ensuremath{N_\text{trk}^\text{offline}}\xspace}
\newcommand{\ncorr}    {\ensuremath{N_\text{trk}^\text{corrected}}\xspace}
\newcommand{\ncorravg} {\ensuremath{\langle \ncorr \rangle}\xspace}
\newcommand {\roots}   {\ensuremath{\sqrt{s}}}
\newcommand{\pp}     {\ensuremath{\Pp\Pp}\xspace}
\newcommand {\PbPb}  {\ensuremath{\text{PbPb}}\xspace}
\newcommand {\pPb}   {\ensuremath{\Pp{}\text{Pb}}\xspace}
\newcommand {\AonA}  {\ensuremath{\text{AA}}\xspace}
\newcommand{\Dzerodecay}{\ensuremath{\PDz\to\PKm\PGpp}\xspace}
\newcommand{\fpromp}{\ensuremath{f^{\text{prompt}}}\xspace}
\ifthenelse{\boolean{cms@external}}{\providecommand{\cmsLeft}{upper\xspace}}{\providecommand{\cmsLeft}{left\xspace}}
\ifthenelse{\boolean{cms@external}}{\providecommand{\cmsRight}{lower\xspace}}{\providecommand{\cmsRight}{right\xspace}}
\newlength{\cmsFigWidth}
\ifthenelse{\boolean{cms@external}}{\setlength{\cmsFigWidth}{0.45\textwidth}}{\setlength{\cmsFigWidth}{0.9\textwidth}}

\DeclareRobustCommand{\PAGLpc}{{\HepParticle{\PGL}{c}{-}}\Xspace} 
\cmsNoteHeader{HIN-21-016}
\title{Multiplicity dependence of charm baryon and charm meson production in \texorpdfstring{\pPb}{pPb} collisions at \texorpdfstring{$\sqrtsNN = 8.16\TeV$}{sqrt(s[NN]) = 8.16 TeV}}

\author*[rice]{Yousen Zhang}

\date{\today}

\abstract{Measurements of the production yields of charm baryons (\PGLpc) and charm mesons (\PDz) in proton-lead collisions at a nucleon-nucleon center-of-mass energy of 8.16\TeV are presented. The data were collected in 2016 with the CMS experiment and correspond to an integrated luminosity of 186 \nbinv. The \PGLpc baryon is reconstructed from the decay channel $\PGLpc \to \PKzS\Pp$,  while the \PDz meson is reconstructed via $\Dzerodecay$. The \PGLpc baryon and \PDz meson yields are extracted in several charged-particle multiplicity classes.  No strong multiplicity dependence of the \PGLpc-to-\PDz yield ratio is observed, unlike the observed strange baryon to strange meson yield ratio of \PGL/\PAGL to \PKzS,  which shows a strong multiplicity dependence. This observation indicates different mechanisms for the multiplicity evolution of hadronization processes for charm and strange quarks and provides new constraints to the understanding of heavy flavor production and collectivity in small collision systems.}

\hypersetup{%
pdfauthor={CMS Collaboration},%
pdftitle={Multiplicity dependence of charm baryon and meson production in pPb collisions at sqrt(s[NN]) = 8.16 TeV},%
pdfsubject={CMS, hadronization},%
pdfkeywords={CMS, hadronization}} %

\maketitle
\section{Introduction}
The quark-gluon plasma (QGP), a deconfined phase of quarks and gluons, is
formed in ultra-relativistic heavy ion collisions. The
collective flow motion (``collectivity'') of this strongly coupled medium in nucleus-nucleus (\AonA)
collisions has been observed at the CERN LHC~\cite{Aamodt:2010pa,Chatrchyan:2012ta,ALICE:2022wpn, CMS:2024krd}
and the BNL RHIC~\cite{Ackermann:2000tr,Adler:2003kt,Adams:2005ph,Alver:2008gk}.
The observed azimuthal anisotropies of the
particles emitted~\cite{Alver:2009id,Abelev:2009af,Chatrchyan:2011eka,Chatrchyan:2012wg,ATLAS:2012at,CMS:2013bza}, such as the ``elliptic flow,''
result from strong interactions between constituents inside the medium that initially has an asymmetric geometrical shape and
can be well described by
hydrodynamic models~\cite{Ollitrault:1992bk,Heinz:2013th,Gale:2013da}.
Similar observations of a common collective signature for charm mesons~\cite{ALICE:2017pbx,Sirunyan:2018toe,CMS:2020bnz} 
and bottom mesons~\cite{ALICE:2020hdw,ATLAS:2020yxw,CMS:2022vfn} 
indicate that heavy quarks may also be strongly coupled to the QGP,
even though their interactions with the medium are expected to be much less frequent than for light quarks.

In recent years, unexpected strong collective flow signals were also observed
in small systems such as proton-proton (\pp)~\cite{Khachatryan:2010gv,Aad:2015gqa,Khachatryan:2015lva,Khachatryan:2016txc} 
and proton-lead (\pPb)~\cite{CMS:2012qk,alice:2012qe,Aad:2012gla,Chatrchyan:2013nka,Aaij:2015qcq,ABELEV:2013wsa,Khachatryan:2014jra,Khachatryan:2015waa,Aaboud:2017acw,Aaboud:2017blb,PHENIX:2018lia} collisions.
The elliptic flow for prompt \PDz~\cite{Sirunyan:2018toe,CMS:2020qul}
and \PJGy~\cite{Acharya:2017tfn,Sirunyan:2018kiz} mesons in \pPb collisions
was found to be comparable to that of light-flavor hadrons at a
given particle transverse momentum (\pt). On the other hand, no evidence for
bottom hadron collective flow has been
observed yet in small systems, such as \pPb collisions~\cite{Aad:2019aol,CMS:2020qul}.
Although effects of final-state partonic rescatterings in a dense medium can successfully
explain a large variety of measurements for light quarks (see reviews in Refs.~\cite{Dusling:2015gta,Nagle:2018nvi}),
they alone cannot describe the large flow for charm quarks~\cite{Du:2018wsj}.
Meanwhile, calculations based on initial-state effects, such as gluon saturation models, can successfully
describe experimental measurements for several charm and bottom mesons~\cite{Zhang:2019dth}.

The study of heavy flavor meson and baryon production yields provides a powerful means 
to further elucidate the origin of heavy flavor collectivity in small systems, as
hadronization processes are also influenced by the presence of final-state rescattering effects.
The abundant deconfined quarks in the QGP can coalesce into bound states before 
fragmenting to low-energy hadrons,
resulting in a different modification of the yields for baryons and mesons~\cite{Fries:2003vb,Fries:2003kq}.
This parton coalescence effect is expected to be
stronger with increasing system size and can enhance the baryon-to-meson yield ratio
at an intermediate hadron \pt range (e.g., 2--6\GeV)~\cite{Plumari:2017ntm,He:2019vgs}.
Models in \AonA collisions incorporating parton coalescence have been able to describe measurements
of heavy-flavor hadrons, such as the \PGLpc-to-\PDz production yield ratio,
at both RHIC and the LHC~\cite{STAR:2019ank,CMS:2019uws,ALICE:2021bib}.
Because of its dependence on final-state effects, the study of
this baryon-to-meson enhancement effect
may also contribute to our understanding of the origin of collectivity in small systems.

Measurements of the charged-particle multiplicity
dependence of the baryon-to-meson ratio for strange particles
show an increasing trend from low to
high multiplicity in the intermediate hadron \pt range~\cite{CMS:2016zzh},
which is qualitatively consistent with models including final-state effects.  
Therefore, an enhancement of charm baryon-to-meson yield ratios should be expected for high multiplicity events,
if final state effects are the origin of heavy flavor quark collectivity. 
For inclusive \pp and \pPb collisions,  models assuming
the presence of a QGP~\cite{Li:2017zuj,Minissale:2020bif} are in agreement
with measurements of
\PGLpc baryon yields ~\cite{ALICE:2020wfu,ALICE:2020wla,ALICE:2022exq}
when coalescence effects are incorporated.
Over the limited range of charged multiplicities in pp collisions,
the statistical hadronization model is able to describe the multiplicity dependence of
the \PGLpc-to-\PDz yield ratio as a function of \pt~\cite{ALICE:2021npz,He:2019tik}.

In this Letter,  the \pt spectra of the \PGLpc and \PDz particles and their ratios are measured 
in the range of 2--10\GeV in \pPb collisions at a nucleon-nucleon 
center-of-mass energy $\sqrtsNN = 8.16\TeV$, for the first time in different multiplicity intervals.
Tabulated results are provided in the HEPData record for this analysis~\cite{hepdata}.

\section{The CMS detector}\label{exp-setup}
The central feature of the CMS apparatus is a superconducting solenoid
of 6\unit{m} internal diameter, providing a magnetic field of 3.8\unit{T}.
Within the solenoid volume are a silicon pixel and strip tracker,
a lead tungstate crystal electromagnetic calorimeter (ECAL),
and a brass and scintillator hadron calorimeter (HCAL),
each composed of a barrel and two endcap sections. 
Hadron forward (HF) calorimeters~\cite{Bayatian:2006jz}, made of steel and quartz fibers, 
extend the pseudorapidity ($\eta=-\ln(\tan(\theta/2))$, where the polar 
angle $\theta$ is defined relative to the counterclockwise beam) coverage provided by the barrel and endcap detectors.
Muons are measured in gas-ionization detectors embedded
in the steel flux-return yoke outside the solenoid. A more detailed
description of the CMS detector, together with a definition of the
coordinate system used and the relevant kinematic variables, can be found in Ref.~\cite{CMS:2008xjf}.

The silicon tracker measures charged particles within the range 
$\abs{\eta_{\text{lab}}} < 2.5$ in the laboratory frame.
During the LHC running period when the
data used in this Letter were recorded, the silicon tracker
consisted of 1440 silicon pixel and 15\,148 silicon strip detector
modules. For particles of $1 < \pt < 10\GeV$ and
$\abs{\eta_{\text{lab}}} < 1.4$, the track resolutions are typically
1.5\% in \pt and 25--90 (45--150)\mum in the transverse (longitudinal)
impact parameter~\cite{CMS:2014pgm}.

The event samples are collected by the CMS experiment with a
two-level trigger
system~\cite{CMS:2020cmk,CMS:2016ngn}. At \Lone,
events are selected by custom hardware processors, while the high-level
trigger (HLT) uses fast versions of the offline software. The \pPb data
used in this analysis were collected in 2016 at $\sqrtsNN = 8.16\TeV$
and correspond to an integrated luminosity of
$186.0\pm6.5\nbinv$~\cite{CMS-PAS-LUM-17-002}.
The beam energies were 6.5\TeV
for the protons and 2.56\TeV per nucleon for the lead nuclei. Because
of the asymmetric beam conditions, the midrapidity range of $\abs{y_{\text{lab}}}<1$,
which is used for particle selections,
corresponds to the rapidity range $-1.46 < y_{\text{cm}} < 0.54$
in the nucleon-nucleon center-of-mass frame, with positive rapidity corresponding
to the proton beam direction. 

The event reconstruction, event
selections, and trigger requirements for minimum bias and high-multiplicity
events are identical to those described in
Refs.~\cite{Khachatryan:2016txc,Sirunyan:2017quh,Sirunyan:2017uyl}.
The minimum bias (MB) 8.16\TeV \pPb events are triggered by requiring energy
deposits above 1\GeV in at least one of the two HF calorimeters and
the presence of at least one track with $\pt > 0.4\GeV$ reconstructed using hits from the pixel tracker only.
To collect a large sample of high-multiplicity \pPb collisions, a dedicated
trigger is implemented. At \Lone, the total number of ECAL+HCAL towers having 
deposited energy above a threshold of 0.5\GeV in transverse energy (\ET)
is required to be greater than 150.
As part of the HLT trigger, the track reconstruction is performed online with
the identical reconstruction algorithm used offline~\cite{CMS:2014pgm}.
For each event selected at \Lone, the reconstructed vertex with the highest number 
of associated tracks is selected as the primary vertex (PV) at the HLT.
The number of online tracks with $\abs{\eta}<2.4$, $\pt > 0.4\GeV$, and a distance of closest
approach less than 0.12\unit{cm} along the beam axis to the PV is 
determined for each event and is required to exceed 185.
The events are required to contain a PV within 15\unit{cm} of 
the nominal interaction point along the beam axis and 0.2\unit{cm} in the
transverse direction. The integrated luminosity sampled by the minimum bias (high-multiplicity)
trigger is 4.22 (97.8) \nbinv.

\section{Data analysis}
Similar to previous measurements~\cite{CMS:2012qk,Chatrchyan:2013nka,Khachatryan:2014jra,Khachatryan:2015waa}, 
hadronic events are selected by requiring the presence of at least one energy deposit greater than 
3\GeV in each of the two HF calorimeters.  The \pPb data are analyzed in several charged-particle multiplicity classes, 
and defined according to the number of selected offline
tracks (\noff) with $\abs{\eta_{\text{lab}}}<2.4$ and $\pt > 0.4\GeV$ that
originate from the PV~\cite{CMS:2014pgm,Khachatryan:2016txc}.
If more than one vertex is found in an event,
the vertex associated with the highest number of offline tracks is
considered as the PV.  During the data taking, the average number of collisions per bunch crossing (pileup) varied from 0.10 to 0.25.
The average number of tracks in each \noff range before and after correcting
for inefficiencies and misidentification rates
in the tracking system is denoted as $\langle \noff \rangle$ and $\ncorravg$, respectively. The statistical uncertainty in
$\ncorravg$ has been found to be negligible, while the systematic
uncertainty is 4\% resulting from the uncertainty of tracking efficiency 
estimated based on simulated events.
The $\langle \noff \rangle$ and $\ncorravg$ values in each \noff range
are summarized in Table~\ref{tab:ntrk-map}; for minimum bias \pPb events, these values are 50 and $59.5\pm0.5$, respectively.
To optimize the signal significance and efficiency,
a sample of simulated signal events is generated using hard processes in \PYTHIA 8.209~\cite{Sjostrand:2014zea}, using the CUETP8M1 tune~\cite{CMS:2019csb}.  These signal events are embedded into \pPb events generated with the {\textsc{EPOS LHC}\xspace} model~\cite{Pierog:2013ria}.
The response of the CMS detector for generated events is simulated using the \GEANTfour~\cite{GEANT4} toolkit.
The decay kinematics for \PGLpc and \PDz particles are simulated using \EVTGEN~\cite{Lange:2001uf}. 

\begin{table}[htb]
  \centering
  \topcaption{The average multiplicity before (and after) corrections,
    $\langle \noff\rangle$
    ($\ncorravg$) with track $\pt>0.4\GeV$ and $\abs{\eta_{\text{lab}}}<2.4$
    in each multiplicity interval.
    The uncertainties reported for \ncorravg are systematic uncertainties,
    as the statistical uncertainties have been found to be negligible.
  }
  \begin{tabular}{lcc}
    \hline \noalign{\smallskip}
    Multiplicity interval (\noff) & $\langle \noff \rangle$ & $\ncorravg$ \\
    \hline \noalign{\smallskip}
    $[2, 35)$ & $16.4$ & $20.0 \pm 0.8$\\
    $[35, 60)$ & $46.3$ & $56.4 \pm 2.3$\\
    $[60, 90)$ & $73.0$ & $88.7 \pm 3.5$\\
    $[90, 120)$ & $102$ & $124 \pm 5$\\
    $[120, 185)$ & $140$ & $170 \pm 7$\\
    $[185, 250)$ & $202$ & $245 \pm 10$\\
    \hline
  \end{tabular}
\label{tab:ntrk-map}
\end{table}

The \PGLpc baryons are reconstructed
using the decay channel $\PGLpc \to \PKzS\Pp $
with the branching fraction of $(1.59\pm0.07)\%$~\cite{ParticleDataGroup:2024cfk}, where
\PKzS mesons are reconstructed via the $\PKzS\to \PGpp \PGpm$ decays with the
branching fraction of $(69.20\pm0.05)\%$~\cite{ParticleDataGroup:2024cfk}.
Throughout this Letter, charge-conjugate modes are implied unless explicitly noted otherwise.
The decay vertex for each \PKzS meson is reconstructed by fitting a common
vertex with two oppositely charged particles inside the CMS tracking detector using a
Kalman filter. Selections based on topological information are also
applied to optimize the \PKzS signal yield significance (over the non-signal hypothesis) of these signals, following the same
procedures as Ref.~\cite{Sirunyan:2018toe}.
The invariant mass of each {\PKzS} candidate ($M_{\PGpp\PGpm}$)
is required to be within $\abs{M_{\PGpp\PGpm}-M_{\PKzS}}<0.02\GeV$, 
where $M_{\PKzS}$ is the \PKzS meson mass~\cite{ParticleDataGroup:2024cfk}.
The secondary decay vertices of the \PGLpc baryon are
reconstructed by combining a  \PKzS candidate with another charged track in the event.
For the \PGLpc reconstruction, each \PKzS candidate is treated as a particle with its kinematic parameters and covariance matrix taken from the \PKzS vertex fit. 
In this step, the \PKzS mass is fixed to the nominal PDG value~\cite{ParticleDataGroup:2024cfk}, rather than using the reconstructed $\PGpp\PGpm$ invariant mass. 
This improves the mass resolution of the \PGLpc candidate.
The \PGLpc candidates (including signals and backgrounds) with invariant mass falling 
within $\abs{M_{\PKzS\Pp} - M_{\PGLpc}}<0.2\GeV$,
are considered in this analysis, where $M_{\PGLpc}$ is the
mass of the \PGLpc baryon~\cite{ParticleDataGroup:2024cfk}.
A multilayer perceptron (MLP) is trained with the Toolkit for 
Multivariate Data Analysis (\textsc{tmva}) package~\cite{Hocker:2007ht} using simulated signal events to
suppress the combinatorial background. The training variables are:
the candidate proton track momentum and $\eta$; the energy loss of the candidate proton track per unit
length inside the tracker normalized by the expected mean value for the proton at the corresponding
momentum; and the cosine of the pointing angle defined as the angle in three dimensions
between the vector connecting the PV with the secondary decay vertex of \PGLpc baryons and
the momentum of the \PGLpc candidate. The energy loss per unit length for each track
is calculated with the ``generalized mean'' of grade--2, as described in Ref.~\cite{Giammanco:2007zz}.

Events populating the sideband region, defined as $0.06<\abs{M_{\PKzS\Pp}-M_{\PGLpc}}<0.11\GeV$ 
that is at least four standard deviations away from the peak of the signal distribution, 
are used as background samples in the TMVA training.
The optimization of the \PGLpc signal significance is performed separately 
for event classes $2 \leq \noff < 185$ (from the minimum bias trigger) and $185 \leq \noff < 250$ (from the high-multiplicity trigger).
The maximum statistical
significance of these signals after optimization is much larger than 5 standard deviations.

The \PDz candidates are reconstructed
using the decay channel $\PDz\to\PKm\PGpp$ with the branching fraction of 
$(3.947\pm0.030)\%$~\cite{ParticleDataGroup:2024cfk}. The boosted decision tree (BDT) method is used to reject combinatorial backgrounds and to
optimize the  \PDz signal yield significance.
Details about the reconstruction, candidate selection, and correction are reported in Ref.~\cite{CMS:2020qul}.

\begin{figure}[htb!]
  \centering
  \includegraphics[width=0.45\textwidth]{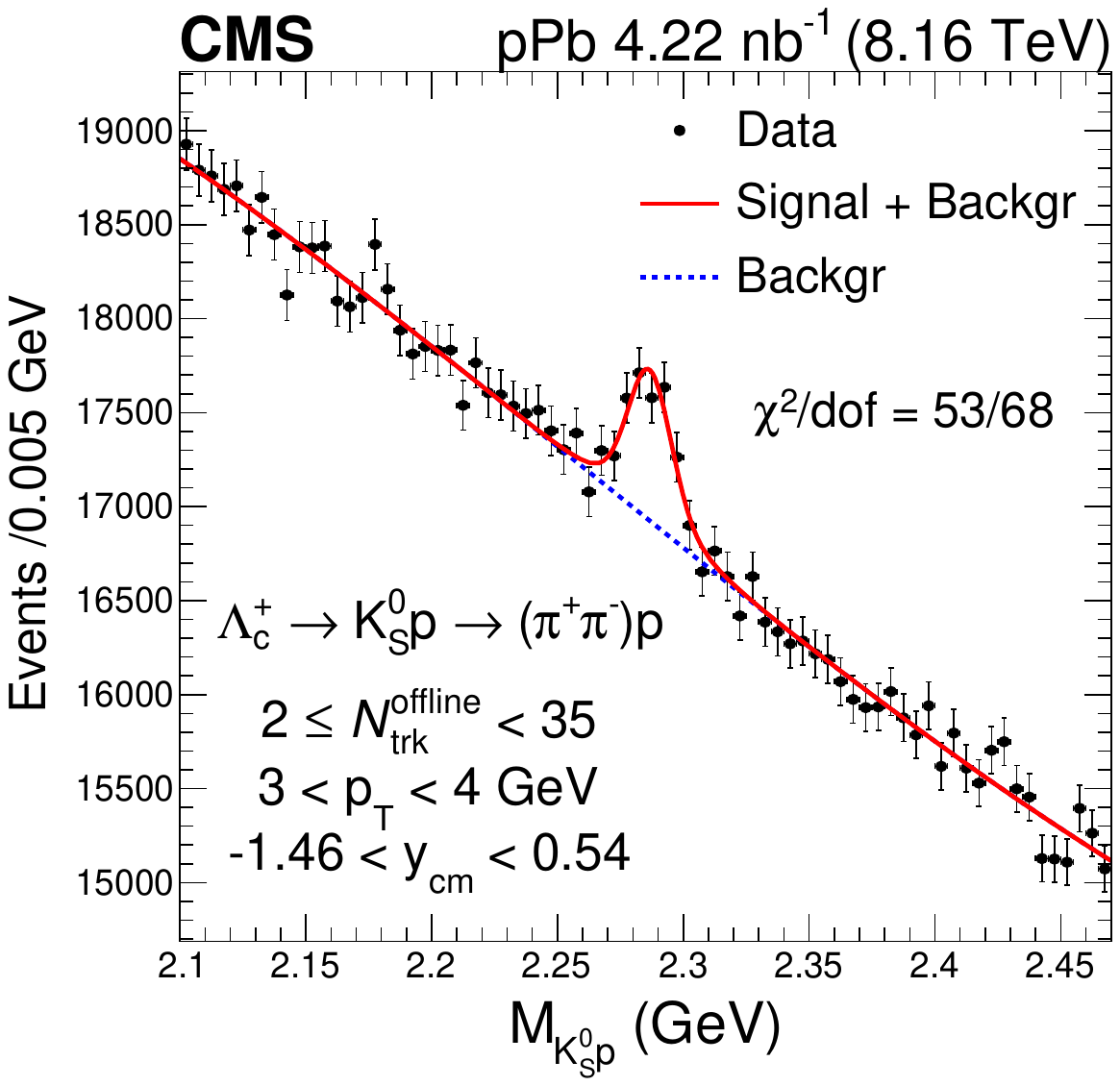}
  \includegraphics[width=0.45\textwidth]{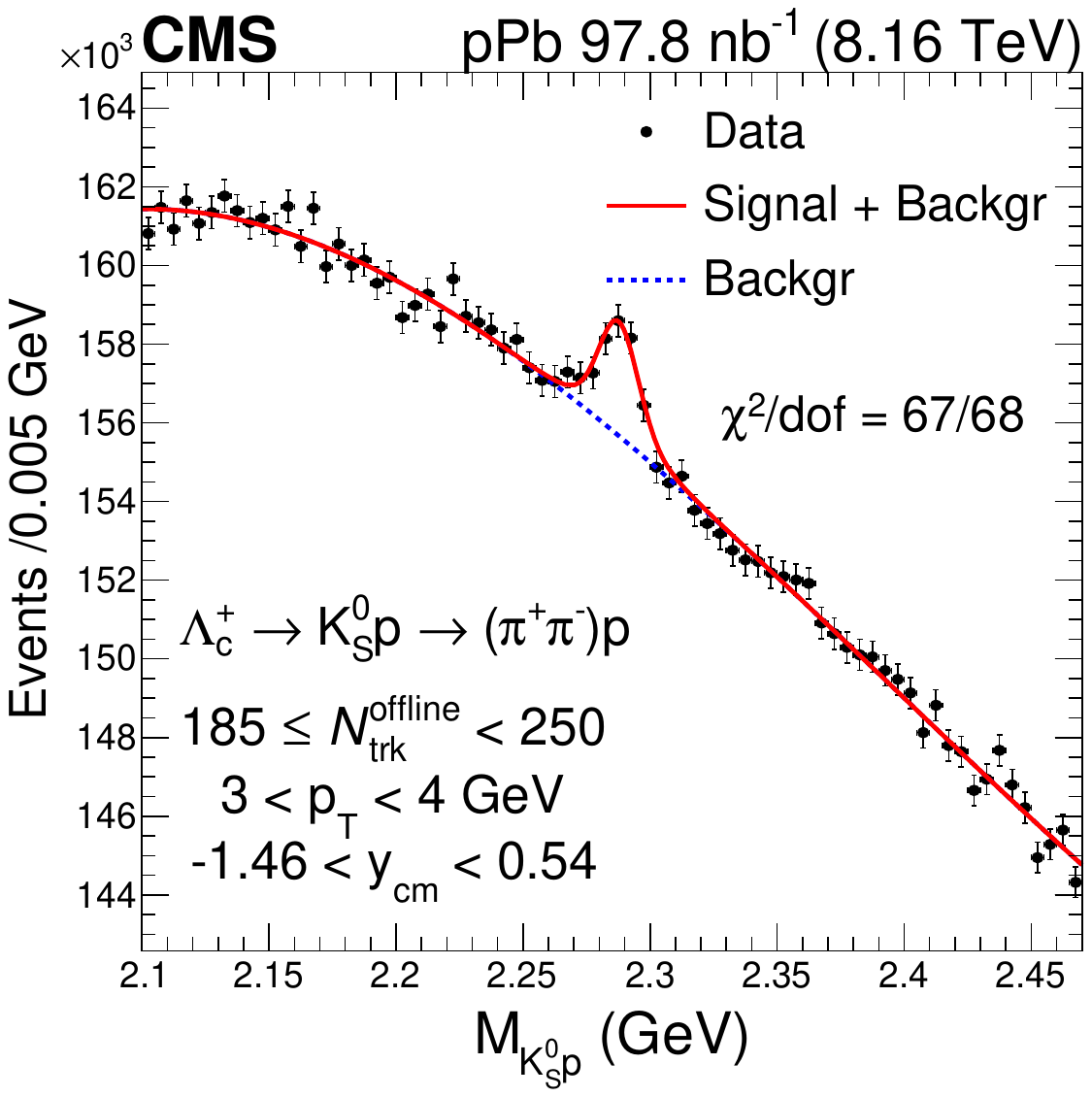}  
  \caption{The fitted mass spectra, along with the goodness-of-fit values, for \PGLpc candidates with
    $3<\pt<4\GeV$ and $-1.46 < y_{\text{cm}} < 0.54$.
    The spectrum for the multiplicity class with $2 \leq \noff<35$
    is shown in the \cmsLeft plot, and the spectrum for $185\leq\noff<250$
    is shown in the \cmsRight plot.}\label{fig:eg-mass}
\end{figure}

The \PGLpc candidate mass spectra are modeled by a double Gaussian function
with a common mean and different widths for the signal contribution and a third-order
Chebyshev polynomial for the background.
The widths of the signal distributions are constrained to
those found in simulation
to improve the fit stability. Figure~\ref{fig:eg-mass} shows
the mass spectra of the \PGLpc candidates for the multiplicity ranges of 
$2\leq\noff<35$ and $185\leq\noff<250$. 
The invariant mass of \PDz candidates is modeled by
several components: a double Gaussian function with a common mean for the \PDz signal; an
additional Gaussian function to describe the invariant mass shape of
\PDz candidates with swapped pion and kaon tracks (``swapped component'');
two Crystal Ball (CB)
functions~\cite{CrystalBallRef} to describe the processes
$\PDz\to\PGpp\PGpm$ and $\PDz\to\PKp\PKm$; and a third-order
polynomial to model the combinatorial background.
Examples of fits to the \PDz
invariant mass are shown in Ref.~\cite{CMS:2020qul}.

The differential yields for \PGLpc and \PDz particles are computed using
\begin{equation}
  \label{eq:yields-per-evt}
  \frac{\rd N}{\rd \pt} =
  \frac{\fpromp N^{\text{sig}}}{2\alpha\epsilon\Delta\pt} \frac{1}{\mathcal{B}},
\end{equation}
where $N^{\text{sig}}$ is the raw yield from the fit to the invariant mass in each \pt and multiplicity bin,
$\alpha$ is the acceptance, $\epsilon$ is the
reconstruction efficiency, $\Delta\pt$ is the width of \pt interval, $\mathcal{B}$ is the branching fraction, and $\fpromp$ is the corresponding prompt component
fraction for \PGLpc and \PDz hadrons, respectively.
Prompt mesons are those directly produced in \pPb collisions or from strong
decays of excited charm hadrons, rather than from decays of bottom hadrons.
The factor of two in the denominator takes into account the inclusion of both charge-conjugate particles in the raw yield.
The acceptance and reconstruction efficiencies are obtained from simulated events in bins of \pt, where
the resolution is much narrower than the bin width.

The fraction $\fpromp$ for \PGLpc events is estimated following a procedure similar to the one described in Ref.~\cite{ALICE:2020wfu}.
The estimation of nonprompt \PGLpc contributions from bottom hadron decays 
considers inputs including the \PGLzb and \PB production cross sections
and the inclusive decay branching fractions of \PGLzb and \PB (\PBp, \PBz) mesons to \PGLpc, \ie,
$\PGLzb \to \PGLpc + \mathrm{X}$ and $\PB \to \PGLpc +\mathrm{X}$.
The \PGLzb production cross section is derived from that for \PB meson production,
computed using the ``fixed order next-to-leading logarithmic'' (FONLL)
perturbative quantum chromodynamics calculation~\cite{Cacciari:2012ny}
and rescaled according to the fragmentation fractions measured
by the LHCb Collaboration~\cite{LHCb:2019fns}.
It is then multiplied by $A = 208$, the atomic number of the lead nucleus, to map it to \pPb collisions. 
Modifications due to nuclear effects are allowed at the level of 25\% in the systematic uncertainties,  as discussed in Section~\ref{sec:syst}.
The $\mathcal{B}$($\PGLzb \to \PGLpc\; X$) value is set to 83\%, as in the \EVTGEN package, and it
includes the contributions from unobserved decay channels~\cite{ALICE:2023wbx}.
The corresponding branching fractions
for the \PB mesons are set to 5\% by summing over all channels
reported in Ref.~\cite{ParticleDataGroup:2024cfk}.
The correlations between the kinematic distributions of \PQb hadrons and
$\PGLpc$ particles are simulated using {\EVTGEN}.
The cross section of nonprompt \PGLpc baryon production
is further corrected for the acceptance and reconstruction efficiency.
The fraction of feed-down contribution is assumed to be multiplicity independent
in the range $2\leq \noff < 185$, as the same candidate selection is applied in all the
multiplicity ranges.
A summary of the nonprompt \PGLpc baryon fraction (1-$\fpromp$), which has little dependence on \pt and \noff within uncertainties, 
is given in Table~\ref{tab:nplamc}.

\begin{table*}[htb]
  \centering
  \renewcommand*{\arraystretch}{1.4}
    \topcaption{Summary table for the fraction of nonprompt \PGLpc baryons (1-$\fpromp$) relative to
    inclusive \PGLpc baryons.
    The fraction is evaluated in different multiplicity ranges.
    Details regarding the systematic uncertainties
    are described in Section~\ref{sec:syst}.
  }
  \begin{tabular}{lcc}
    \hline
    \pt (\GeVns{}) & $2 \leq \noff < 185$ (\%) & $185 \leq \noff < 250 $ (\%) \\
    \hline
    2--3 & $6.8_{-6.8}^{+11}$ & \NA \\
    3--4 & $4.1_{-4.1}^{+5.8}$ & $4.3_{-4.3}^{+6.0}$ \\
    4--5 & $3.3_{-3.3}^{+5.5}$ & $3.4_{-3.4}^{+5.7}$ \\
    5--6 & $3.6_{-2.9}^{+2.9}$ & $3.6_{-3.0}^{+3.0}$ \\
    6--8 & $3.5_{-3.5}^{+3.8}$ & $3.1_{-3.1}^{+3.4}$ \\
    8--10 & $3.9_{-3.9}^{+5.5}$ & $4.3_{-4.3}^{+6.2}$ \\
    \hline
  \end{tabular}
  \label{tab:nplamc}
\end{table*}

The $\fpromp$ fraction for \PDz mesons is obtained
using a template fit method.
The prompt and nonprompt \PDz mesons can be distinguished based on
the distributions of their distance of closest approach (DCA)
with respect to the PV of the event. A two-component fit with DCA templates
from simulated prompt and nonprompt \PDz samples is performed to the DCA distribution
in data. Examples of the template fit to DCA distributions for \PDz
mesons can be found in Ref.~\cite{CMS:2020qul}.

\begin{figure}[htb!]
\centering
\includegraphics[width=0.45\textwidth]{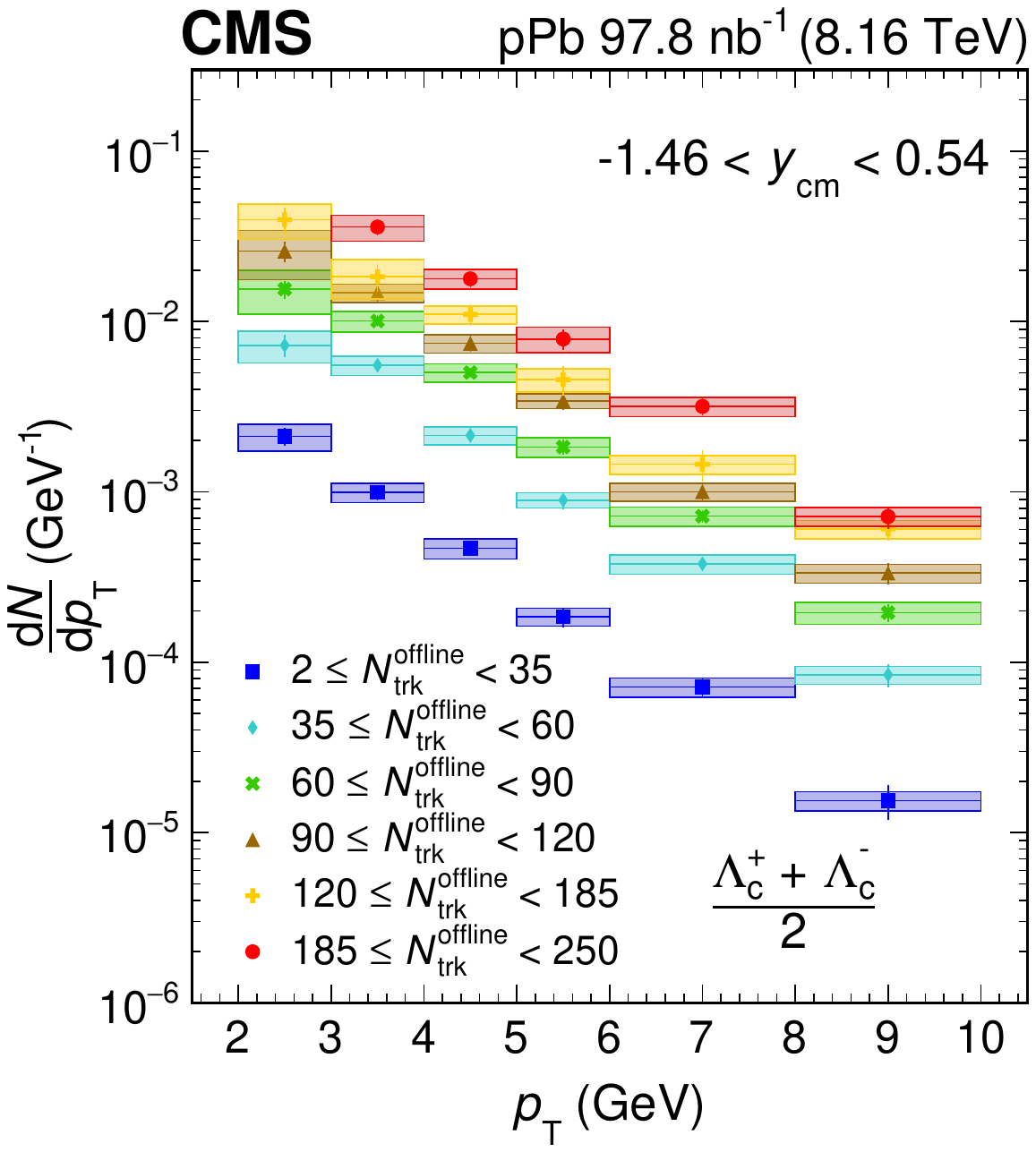}
\includegraphics[width=0.45\textwidth]{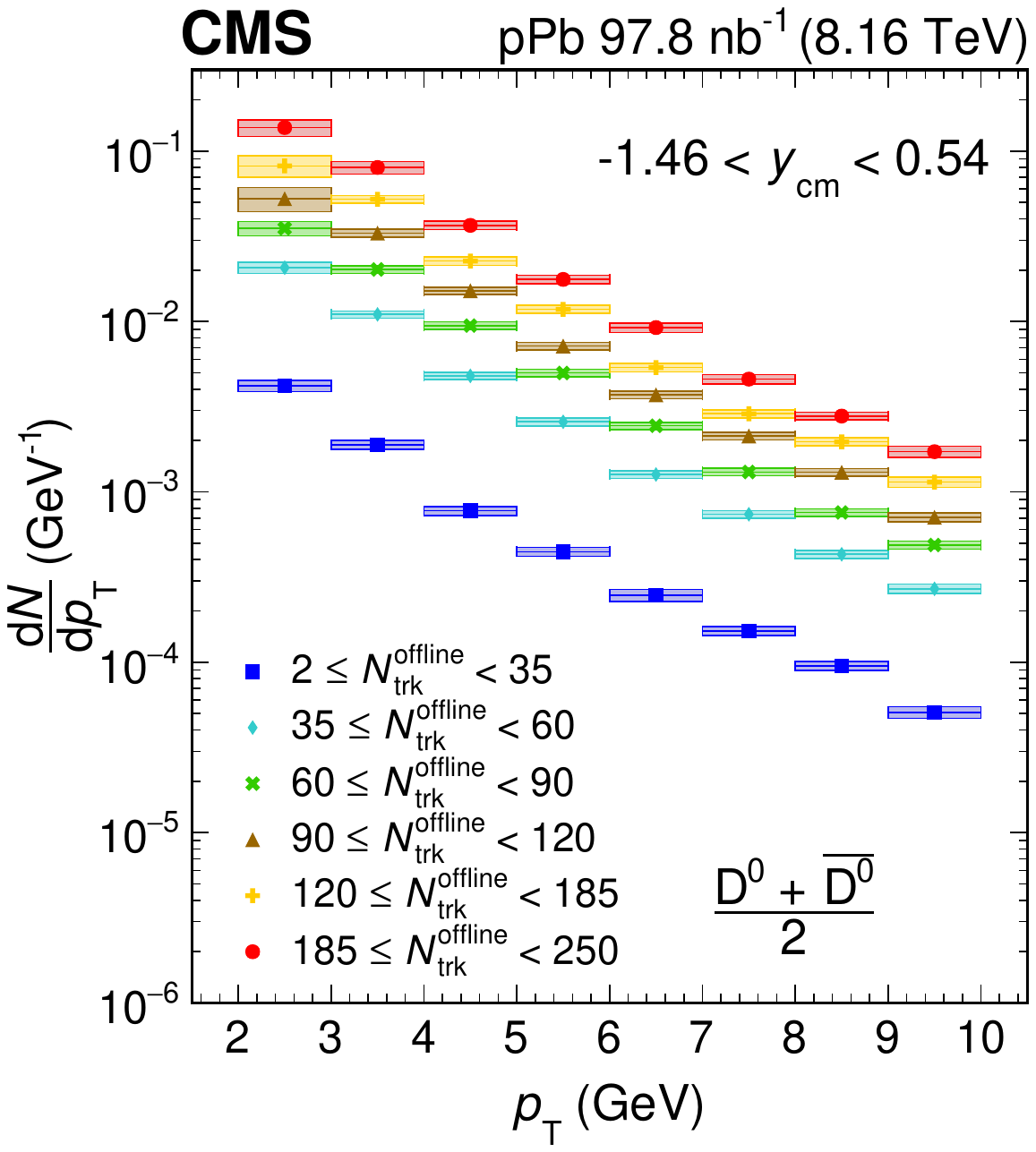}
\caption{Transverse momentum spectra for \PGLpc baryons (\cmsLeft) and \PDz mesons
  (\cmsRight) with $-1.46 < y_{\text{cm}} < 0.54$
  in different multiplicity classes. The vertical bars show the statistical uncertainties while the shaded areas
  represent the systematic uncertainties.}\label{fig:spectra}
\end{figure}

\section{Systematic uncertainties}\label{sec:syst}

The candidate selection uncertainty is estimated by varying the MLP or BDT classifier cut over a wide range, which yields
a relative uncertainty of 2--6\% for \PGLpc baryons and 1--5\% for \PDz mesons, respectively, depending on the \noff and charm hadron \pt.
The uncertainty of the fit procedure is estimated by changing the signal
shape from a double Gaussian to a Gaussian+CB function for the \PGLpc baryon, and via
changing the background function to a second- and fourth-order polynomial
for both \PGLpc and \PDz particles. These variations yield a
relative uncertainty of 6--28\% for \PGLpc baryons and less than 2\% for \PDz mesons.

The uncertainty in feed-down from \PQb hadrons for \PGLpc
is estimated from the following components. The associated systematic
uncertainty from FONLL calculations is evaluated by varying its central values
by its uncertainties~\cite{Cacciari:2012ny}. Another variation is performed by comparing
the FONLL calculations with \PYTHIA.
The value of the Pb-normalized FONLL calculations is also varied by 25\% to account for potential nuclear effects.
No significant nuclear modifications to the \PQb-hadrons production cross sections
in \pPb collisions are observed within $\pm25\%$~\cite{LHCb:2019avm}.
The uncertainty from the LHCb fragmentation fraction measurements is also considered~\cite{LHCb:2019fns}.
The assumption that the feed-down is uniform across multiplicities is
studied using \PDz mesons and assumed to have similar behavior for \PGLpc baryons,
because both nonprompt \PGLpc and \PDz production originates from \PQb hadrons.
The maximum of the difference of $\fpromp$ for \PDz mesons between different multiplicity classes
is taken as an estimate for the systematic uncertainty in the \PGLpc measurements.
The sum in quadrature of these sources yields a total relative uncertainty of 4--12\%
for the feed-down estimation. The uncertainty
related to template fits in DCA for \PDz mesons is estimated by varying the DCA resolution 
of simulated samples to explore the discrepancy between simulated samples and data.
This source yields an uncertainty of less than 15\%.

The systematic uncertainty
from the tracking efficiency is 2.3\% per track~\cite{CMS-DP-2018-050}, 
which is propagated to
7\% for \PGLpc baryon yields and 5\% for \PDz meson yields since the former are
reconstructed from three tracks and the latter are from two.
The uncertainty related to trigger efficiency is found to be negligible,
as the main inefficiency affects mostly very low multiplicity events 
while those containing a heavy flavor hadron production are
strongly biased toward higher average multiplicity within each class. The pileup effects are studied by requiring different minimum distances between 
multiple reconstructed PVs according to the final-state track multiplicity.
The systematic uncertainty due to pileup effects turns out to be 3--6\% (from low to high \noff)
for \PGLpc baryons and $<1$\% for \PDz mesons.
The total systematic uncertainty of the production yields is calculated
by assuming all sources are independent and adding them in quadrature.

The relative systematic uncertainty for the yield ratio
of \PGLpc-to-\PDz is computed by adding in quadrature
the relative \PGLpc and \PDz yield uncertainties including the effects of
candidate selections, fit procedures, \PQb hadron feed-down, and branching fractions.
The uncertainties from tracking efficiency and pileup effects are
taken to be fully correlated between the \PDz and \PGLpc samples, 
resulting in uncertainties of 2.3\% and 2--4\% for the \PGLpc-to-\PDz ratio, respectively.

A summary of systematic uncertainties in measurements of the \PGLpc and \PDz 
differential yields can be found in Table~\ref{tab:syst-sum}. 

\begin{table}[htb]
  \centering
  \topcaption{Summary of relative uncertainties in the differential yields.
  }  
  \begin{tabular}{lcc}
    \hline
    Source & \PGLpc (\%) & \PDz (\%) \\
    \hline
    Selections & 2--6 & 1--5 \\
    Fit & 6--28 & $<$2 \\
    \PQb feed-down & 3--11 & $<$15 \\
    Tracking efficiency & 7 & 5 \\
    Pileup effects & 3--6 & $<$1\\
    Branching fraction & 5 & 1 \\
    \hline
    Total & 10--32 & 5--16 \\
    \hline
  \end{tabular}
\label{tab:syst-sum}
\end{table}

\begin{figure}[htb]
\centering
\includegraphics[width=\cmsFigWidth]{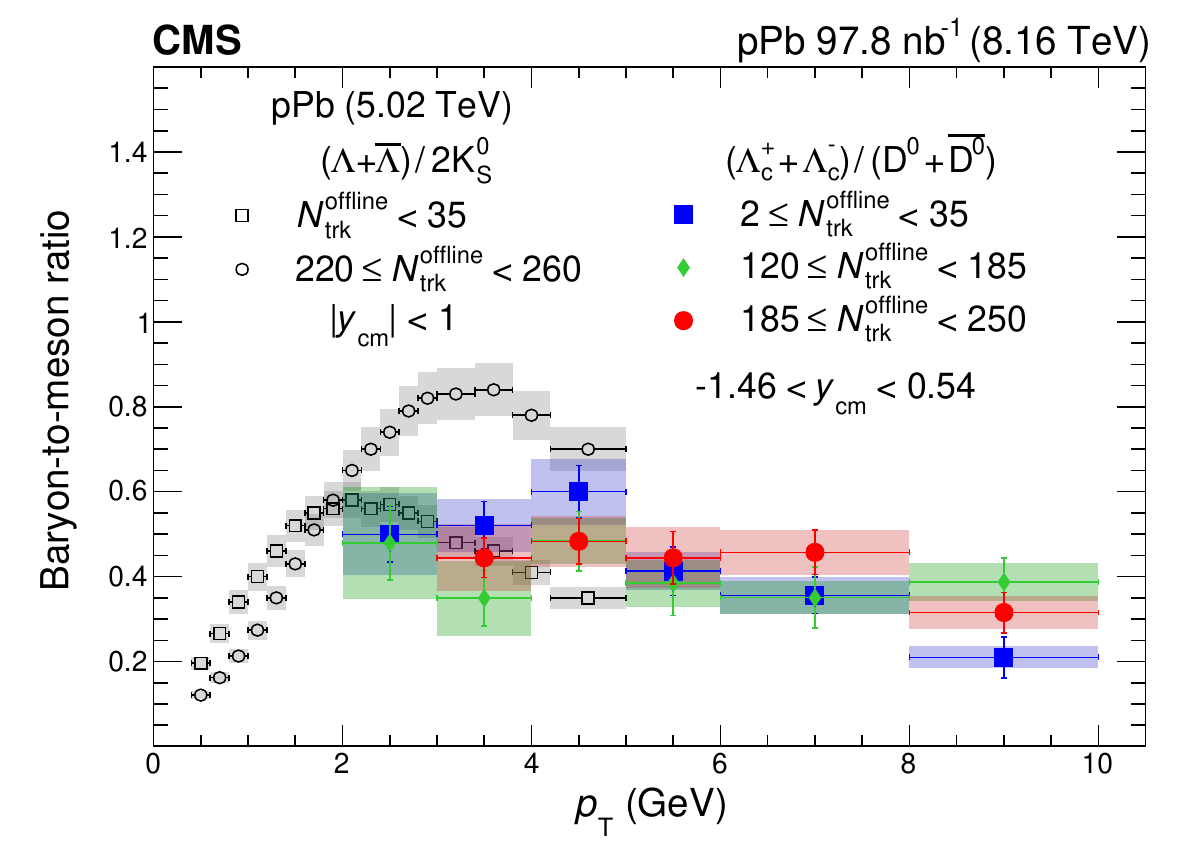}
\caption{Ratio of \PGLpc baryons to \PDz mesons production
  in \pPb collisions at $\sqrtsNN=8.16\TeV$ with
  \( 2 \leq \noff <35\),  \(120 \leq \noff < 185\) and \(185 \leq \noff < 250\). 
  The vertical bars show the statistical uncertainties while the
  shaded areas represent the systematic uncertainties. The grey
  markers are for \PGL baryon to (two times) \PKzS meson production
  at $\sqrtsNN=5.02\TeV$ from Ref.~\cite{CMS:2016zzh}.}\label{fig:ratio-pt}
\end{figure}

\section{Results}\label{sec:results}

The \pt spectra for \PGLpc and \PDz hadrons in each multiplicity class
are shown in Fig.~\ref{fig:spectra}. 
The production of charm hadron yield per event
increases with multiplicity.  The \pt spectra follow approximately an exponential falling trend for all multiplicity classes. 

The corresponding \PGLpc-to-\PDz ratios are presented in Fig~\ref{fig:ratio-pt} for three multiplicity classes.
The \PGLpc-to-\PDz ratio tends to decrease with \pt, both for low- and high-multiplicity events.
The ratios in Fig.~\ref{fig:ratio-pt} are
also consistent with those obtained in inclusive \pPb collisions at $\sqrtsNN=5.02\TeV$
reported in Refs.~\cite{ALICE:2020wfu,ALICE:2022exq}.
No strong
multiplicity dependence is observed within one standard deviation, in contrast
to the enhanced baryon-to-meson ratio for strange hadrons as the 
multiplicity increases~\cite{CMS:2016zzh}.

\begin{figure}[htb]
  \centering
  \includegraphics[width=\cmsFigWidth]{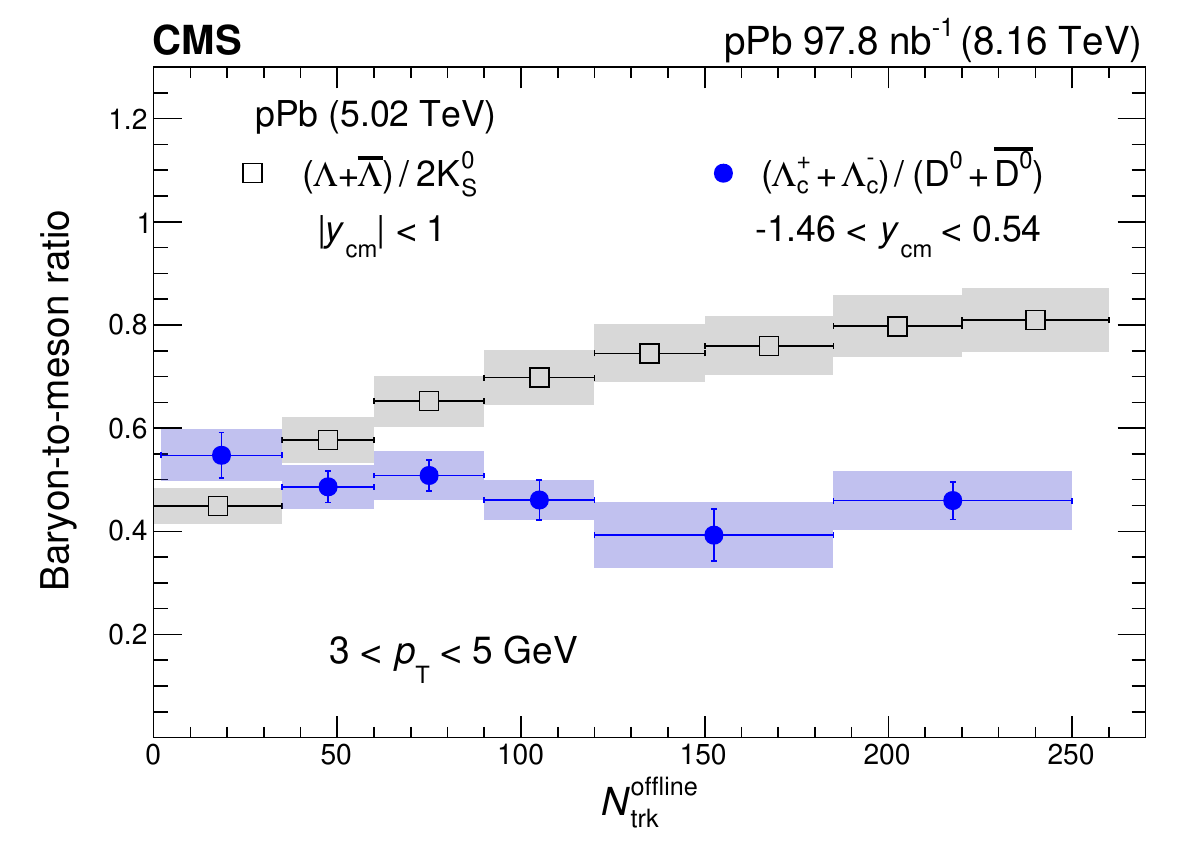}
  \caption{Ratio of \PGLpc baryon to \PDz meson production with
    $-1.46 < y_{\text{cm}} < 0.54$ in \pPb collisions at $\sqrtsNN=8.16\TeV$
    as a function of \noff. The vertical bars show the statistical uncertainties while the shaded
    areas represent the systematic uncertainties. The ratio of
    \PGL baryon to (two times) \PKzS meson production at $\sqrtsNN=5.02\TeV$ obtained from data in 
    Ref.~\cite{CMS:2016zzh} is also shown for comparisons, denoted by
    grey markers.}\label{fig:ratio-ntrk}
\end{figure}

To further investigate the system size dependence, the charm
baryon-to-meson ratio is
also plotted as a function of \noff in Fig.~\ref{fig:ratio-ntrk}.
In contrast to what is observed in the strange-quark sector,
the charm baryon-to-meson ratio is approximately constant
over a large range of multiplicity. This may indicate a possible mass
dependence for the particle hadronization mechanism.
If a coalescence process is present, it may happen earlier
for charm quarks than for light quarks during the system evolution.

The observed independence of the \PGLpc-to-\PDz ratio on the multiplicity might
indicate a difference between heavy and light quarks, even though these particles
show similar elliptic flow signals as a function of multiplicity~\cite{Sirunyan:2018toe,Sirunyan:2018kiz}.
In models of QGP formation, charm quarks are expected to experience kinetic propagation
inside the liquid QGP~\cite{He:2019vgs,Beraudo:2022dpz},
while light quarks are in general thermalized in the hot QGP.
The larger mean free path of heavy quarks compared with light quarks 
in \pPb collisions may lead to less pronounced effects from the dense medium.
On the other hand, a model incorporating an initial-state gluon saturation scenario
without QGP production shows agreement with the elliptic flow of \PDz mesons
in \pPb collisions~\cite{Zhang:2019dth,Zhang:2020ayy}.

\begin{figure}[htb]
  \centering
  \includegraphics[width=\cmsFigWidth]{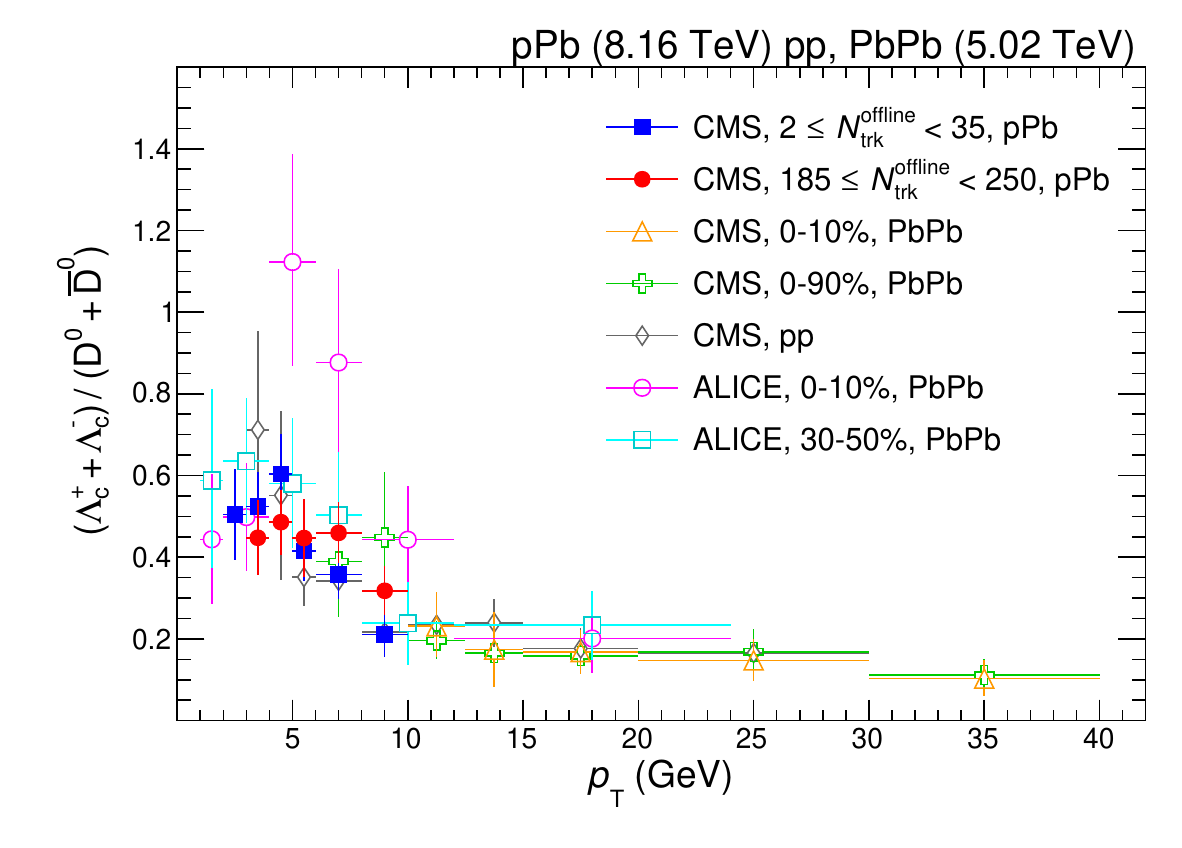}
  \caption{Comparison between the \PGLpc-to-\PDz ratio as a function of
    \pt in \pp, \pPb, and \PbPb (different centrality ranges) collisions
    at the nucleon-nucleon center-of-mass energies of 5.02, 8.16, and 5.02\TeV,
    respectively~\cite{ALICE:2021bib,CMS:2023frs}.
    The ratio is measured with the rapidity ranges $-1.46 < y_{\text{cm}} < 0.54$ in
    \pPb collisions and $\abs{y_{\text{cm}}}<1$ in \pp and \PbPb collisions by the CMS Collaboration~\cite{CMS:2023frs}
    and $\abs{y_{\text{cm}}}<0.5$ by the ALICE Collaboration~\cite{ALICE:2021bib}. The vertical bars indicate the uncertainties. For pPb, the uncertainty is the quadratic sum of statistical and systematic uncertainties; for pp and PbPb, the uncertainties are as reported in Ref.~\cite{ALICE:2021bib} and~\cite{CMS:2023frs}, respectively.    
}\label{fig:ratio-summary}
\end{figure}

The charm baryon-to-meson ratio as a function of \pt in \pPb collisions at 8.16\TeV
is also compared with that in \pp and \PbPb collisions at 5.02\TeV~\cite{ALICE:2021bib,CMS:2023frs}
in Fig.~\ref{fig:ratio-summary}.
The results in the range $2<\pt<10\GeV$ are consistent with inclusive events
in pp collisions and mid-central (\eg, 30--50\% centrality) \PbPb collisions. The
decreasing trend with \pt for the charm baryon-to-meson ratio is also observed in
these systems. 
Further studies of the multiplicity dependence of this ratio
in \pPb collisions for the very high-\pt regime are crucial to find out
whether this convergence is universal across different colliding systems
and event activities.

\section{Summary}

The first measurements of {\PGLpc}
baryon and \PDz meson yields, as well as their yield ratios,
as a function of the charged-particle multiplicity 
in proton-lead (\pPb) collisions are presented. At a nucleon-nucleon center-of-mass energy of 8.16\TeV, 
the \PGLpc baryon is reconstructed using the decay channel $\PGLpc \to \PKzS \Pp$,
while the \PDz meson is reconstructed from $\Dzerodecay$.
The \PGLpc and \PDz  production yields, and the \PGLpc-to-\PDz yield ratio are studied 
as a function of transverse momentum and charged-particle multiplicity.
No strong multiplicity dependence is observed within the experimental uncertainties.
The absence of any significant multiplicity dependence 
of the yield ratio differs strikingly from that for strange hadrons,
which is observed to increase with multiplicity.
The difference between these results
for charm quarks and those for strange quarks
might indicate that the conjectured coalescence processes of heavy quarks happen earlier than
those of strange quarks. 

\begin{acknowledgments}
  We congratulate our colleagues in the CERN accelerator departments for the excellent performance of the LHC and thank the technical and administrative staffs at CERN and at other CMS institutes for their contributions to the success of the CMS effort. In addition, we gratefully acknowledge the computing centers and personnel of the Worldwide LHC Computing Grid and other centers for delivering so effectively the computing infrastructure essential to our analyses. Finally, we acknowledge the enduring support for the construction and operation of the LHC, the CMS detector, and the supporting computing infrastructure provided by the following funding agencies: SC (Armenia), BMBWF and FWF (Austria); FNRS and FWO (Belgium); CNPq, CAPES, FAPERJ, FAPERGS, and FAPESP (Brazil); MES and BNSF (Bulgaria); CERN; CAS, MoST, and NSFC (China); MINCIENCIAS (Colombia); MSES and CSF (Croatia); RIF (Cyprus); SENESCYT (Ecuador); ERC PRG, RVTT3 and MoER TK202 (Estonia); Academy of Finland, MEC, and HIP (Finland); CEA and CNRS/IN2P3 (France); SRNSF (Georgia); BMBF, DFG, and HGF (Germany); GSRI (Greece); NKFIH (Hungary); DAE and DST (India); IPM (Iran); SFI (Ireland); INFN (Italy); MSIP and NRF (Republic of Korea); MES (Latvia); LMTLT (Lithuania); MOE and UM (Malaysia); BUAP, CINVESTAV, CONACYT, LNS, SEP, and UASLP-FAI (Mexico); MOS (Montenegro); MBIE (New Zealand); PAEC (Pakistan); MES and NSC (Poland); FCT (Portugal); MESTD (Serbia); MCIN/AEI and PCTI (Spain); MOSTR (Sri Lanka); Swiss Funding Agencies (Switzerland); MST (Taipei); MHESI and NSTDA (Thailand); TUBITAK and TENMAK (Turkey); NASU (Ukraine); STFC (United Kingdom); DOE and NSF (USA).
   
  \hyphenation{Rachada-pisek} Individuals have received support from the Marie-Curie program and the European Research Council and Horizon 2020 Grant, contract Nos.\ 675440, 724704, 752730, 758316, 765710, 824093, 101115353, 101002207, and COST Action CA16108 (European Union); the Leventis Foundation; the Alfred P.\ Sloan Foundation; the Alexander von Humboldt Foundation; the Science Committee, project no. 22rl-037 (Armenia); the Belgian Federal Science Policy Office; the Fonds pour la Formation \`a la Recherche dans l'Industrie et dans l'Agriculture (FRIA-Belgium); the F.R.S.-FNRS and FWO (Belgium) under the ``Excellence of Science -- EOS" -- be.h project n.\ 30820817; the Beijing Municipal Science \& Technology Commission, No. Z191100007219010 and Fundamental Research Funds for the Central Universities (China); the Ministry of Education, Youth and Sports (MEYS) of the Czech Republic; the Shota Rustaveli National Science Foundation, grant FR-22-985 (Georgia); the Deutsche Forschungsgemeinschaft (DFG), among others, under Germany's Excellence Strategy -- EXC 2121 ``Quantum Universe" -- 390833306, and under project number 400140256 - GRK2497; the Hellenic Foundation for Research and Innovation (HFRI), Project Number 2288 (Greece); the Hungarian Academy of Sciences, the New National Excellence Program - \'UNKP, the NKFIH research grants K 131991, K 133046, K 138136, K 143460, K 143477, K 146913, K 146914, K 147048, 2020-2.2.1-ED-2021-00181, and TKP2021-NKTA-64 (Hungary); the Council of Science and Industrial Research, India; ICSC -- National Research Center for High Performance Computing, Big Data and Quantum Computing and FAIR -- Future Artificial Intelligence Research, funded by the NextGenerationEU program (Italy); the Latvian Council of Science; the Ministry of Education and Science, project no. 2022/WK/14, and the National Science Center, contracts Opus 2021/41/B/ST2/01369 and 2021/43/B/ST2/01552 (Poland); the Funda\c{c}\~ao para a Ci\^encia e a Tecnologia, grant CEECIND/01334/2018 (Portugal); the National Priorities Research Program by Qatar National Research Fund;MCIN/AEI/10.13039/501100011033, ERDF ``a way of making Europe", and the Programa Estatal de Fomento de la Investigaci{\'o}n Cient{\'i}fica y T{\'e}cnica de Excelencia Mar\'{\i}a de Maeztu, grant MDM-2017-0765 and Programa Severo Ochoa del Principado de Asturias (Spain); the Chulalongkorn Academic into Its 2nd Century Project Advancement Project, and the National Science, Research and Innovation Fund via the Program Management Unit for Human Resources \& Institutional Development, Research and Innovation, grant B39G670016 (Thailand); the Kavli Foundation; the Nvidia Corporation; the SuperMicro Corporation; the Welch Foundation, contract C-1845; and the Weston Havens Foundation (USA).
\end{acknowledgments}

\bibliography{auto_generated} \cleardoublepage \appendix\section{The CMS Collaboration \label{app:collab}}\begin{sloppypar}\hyphenpenalty=5000\widowpenalty=500\clubpenalty=5000
\cmsinstitute{Yerevan Physics Institute, Yerevan, Armenia}
{\tolerance=6000
A.~Hayrapetyan, A.~Tumasyan\cmsAuthorMark{1}\cmsorcid{0009-0000-0684-6742}
\par}
\cmsinstitute{Institut f\"{u}r Hochenergiephysik, Vienna, Austria}
{\tolerance=6000
W.~Adam\cmsorcid{0000-0001-9099-4341}, J.W.~Andrejkovic, T.~Bergauer\cmsorcid{0000-0002-5786-0293}, S.~Chatterjee\cmsorcid{0000-0003-2660-0349}, K.~Damanakis\cmsorcid{0000-0001-5389-2872}, M.~Dragicevic\cmsorcid{0000-0003-1967-6783}, P.S.~Hussain\cmsorcid{0000-0002-4825-5278}, M.~Jeitler\cmsAuthorMark{2}\cmsorcid{0000-0002-5141-9560}, N.~Krammer\cmsorcid{0000-0002-0548-0985}, A.~Li\cmsorcid{0000-0002-4547-116X}, D.~Liko\cmsorcid{0000-0002-3380-473X}, I.~Mikulec\cmsorcid{0000-0003-0385-2746}, J.~Schieck\cmsAuthorMark{2}\cmsorcid{0000-0002-1058-8093}, R.~Sch\"{o}fbeck\cmsorcid{0000-0002-2332-8784}, D.~Schwarz\cmsorcid{0000-0002-3821-7331}, M.~Sonawane\cmsorcid{0000-0003-0510-7010}, S.~Templ\cmsorcid{0000-0003-3137-5692}, W.~Waltenberger\cmsorcid{0000-0002-6215-7228}, C.-E.~Wulz\cmsAuthorMark{2}\cmsorcid{0000-0001-9226-5812}
% ...
\par}
\vskip\cmsinstskip
\dag:~Deceased\\
$^{1}$Also at Yerevan State University, Yerevan, Armenia\\
$^{2}$Also at TU Wien, Vienna, Austria\\
% ...
\end{sloppypar}
\end{document}